\documentclass[a4paper,12pt]{article}
\usepackage{amssymb,amsfonts,enumerate}
\usepackage{graphicx,array}

\newlength{\defbaselineskip}
\begin{document}

\title{In Appreciation of Abner Shimony}
\author{Gregg Jaeger\\
             	Quantum Communication and Measurement Laboratory,\\ 
	         Department of Electrical and Computer Engineering\\
             	and Division of Natural Science and Mathematics,\\ Boston University, Boston, MA \\
}\maketitle

Abner Shimony was an exceptional human being and a remarkably lucid and penetrating thinker whose work centered on some of the most significant physical and philosophical questions of his era at their nexus. He approached these questions with an open, agile and critical mind, something quickly evident to anyone who had the privilege of conversing with him.  His choice of problems to pursue, which for the most part involved epistemology and the relationships between mind, matter and space-time, was visionary. His standards, brought continually to bear, were sterling. He was also a charitable man, with great concern for the well being of others, and was politically active, particularly in promoting international peace in the 1980s.

The work of Abner Shimony was driven in no respect by fashion, but by curiousity coupled with a sense of intellectual urgency. He possessed a rarely found range of sophisticated tools necessary for attacking interdisciplinary questions in the foundations of the human conception of the world, arguably one of the few with such capacity since the time of Descartes.
He also saw morality in intellectual endeavors as of prime significance, as evidenced in his view that the moral character of his friend John Bell  was primarily responsible for his discovery of Bell's Theorem. This character was something the two friends shared, Shimony considering Bell one of the most rigorously honest thinkers of all time.\footnote{Together with Roman Jackiw, Shimony wrote an extended article documenting the work and life of John Bell that reflected this view. R. Jackiw and A. Shimony, ``In appreciation of the depth and breadth of John Bell's physics,'' {\em Physics in Perspective}, {\bf 4}, 78 (2002).} Shimony possessed a singular manner of speaking and of precise English usage finely tuned to his task, communicating ideas with the highest fidelity; one who had the privilege of hearing him speak or conversing with him can readily hear this voice in his own mind's ear when considering one of his texts.\footnote{Shimony's English skills, he was quick to point out, were honed with the assistance of his first wife Annemarie Anrod Shimony, a fellow academic whom he met and married during their studies and who supported him steadfastly during his two Ph.D. studies, first in philosophy and later in physics. His English was also well supported by his excellent knowledge of Latin and his experience in French.} One of the phrases one found him using in conversational mode is the phrase ``knowing what one is talking about''; he continually emphasized the importance of being very aware that when one uses terminology, one must take exceptional care that it is appropriate to the context at hand.\footnote{Such mistakes have cost the community working in foundational problems of physics valuable time, as he pointed out, cf. https://www.aip.org/history-programs/niels-bohr-library/oral-histories/25643.} He was also quick to point out the original source for any idea, never allowing undeserved credit to be given if it might appear in the slightest to be given to the incorrect person, particularly when that person might be he himself.

Shimony first graduated {\em summa cum laude} in mathematics and philosophy from Yale, where he entered as an undergraduate as a 16-year-old interested in the sciences and already an enthusiastic supporter of the evolutionary point of view, from which he never wavered. In his period of graduate studies he first moved on to the University of Chicago, where he was advised in philosophy by Rudolf Carnap and then returned to the graduate philosophy program at Yale, completing his Ph.D. there in 1953 after which, having also been a student of mathematics, he served two years in the US Army Signals Corps of Engineers, during which tenure he performed many related calculations and became familiar with the communications theory of Claude Shannon. After this, in 1955, Shimony entered Princeton to study physics, which he did under Eugene Wigner, writing his Ph.D. thesis in statistical mechanics. 

Upon that final of his graduations, Shimony accepted a post at MIT in the Department of Humanities, returning to investigations in epistemology and beginning what was to become a concentrated focus on the foundations of quantum mechanics, all the while keeping in touch with and supported by Wigner in relation to work in the latter. This was natural for him, as he saw the two subjects as deeply interrelated. In 1968, as would later a number of excellent scholars from MIT, he moved to Boston University,  primarily because there he was offered jointly two positions, one in the Physics Department and one in the Philosophy Department, which he would come to hold for twenty-six years.\footnote{It was Boston University where I myself came for graduate studies in physics, specifically to study under Abner Shimony for reasons evident in this article, becoming his final Ph.D. student of physics.} During this tenure, he also taught as a visitor at the Sorbonne, the University of Geneva and ETH, and received (in 1996) the Lakatos Award for his outstanding contributions to the philosophy of science, among other accumulated honors and titles.  

Undoubtedly, most of the discussions of Abner Shimony's work in future will center on his enormous contribution to the investigation of the significance to physics, both theoretical and experimental, of quantum entanglement, the characteristic of physical entities on which his work came to concentrate. Entanglement is indeed the greatest thread running through the fabric of his work, well represented in the result---appearing in a paper published together with John Clauser and Shimony's once student and afterward long-time collaborator Michael Horne, as well as the then Harvard Ph. D. candidate Richard Holt---now known as the Clauser--Horne--Shimony--Holt inequality. This result is a directly empirically testable form of the now famous but then less well known Bell inqualities. He also made a signficant analytical contribution to reinforcing and bringing out the full empirical significance of another testable mathematical relation involving entanglement, one produced by his long-time collaborators Horne, (Daniel) Greenberger and (Anton) Zeilinger, together known as GHZ. Shimony's early work on entanglement can be found in the two-volume collection of his early writings, {\em The Search for a Naturalistic World View} (Cambridge University Press, 1993).\footnote{Shimony's other works on entanglement and other matters that were to follow are presently uncollected, but are unlikely to remain so for long; for the time being, one can look to a later volume of papers dedicated to him stemming from a conference held in his honor at the Perimeter Institute hosted by PI, edited by two other of his former students, Wayne Myrvold and Joy Christian who were residing near the PI at the time, entitled {\em Quantum Reality, Relativistic Causality, and Closing the Epistemic Circle}, contributed by researchers from a range of fields and specializations in physics and philosophy.}

Also significant for physics is his pioneering development of the {\em quantification} of entanglement in his 1995 article ``Degree of Entanglement,'' perhaps the very first work to address this question, developed from a set of criteria, neither {\em ad hoc} nor stretched to fit any preconceived notion. This article included among the workable measures it identified not only the later very popular and standard approach of measuring entanglement entropically that has been favored by those in computer-scientifically directed investigations in which it is quantified via a sort of entropy, but also the more physical (though in case of infinite dimensionality, also problematic) approach in which it is considered geometrically, within the corresponding powerful and elegant tradition of physics.\footnote{Just beforehand, Shimony, Horne and I had arrived at an empirical measure of entanglement, a non-trivially corrected joint-system interference visibility, which Artur Ekert would subsequently call the ``visibility of entanglement,'' and demonstrated a complementarity between it and single-system interference visibility for two-level systems.} Now, however, rather than on this work, let us consider instead Abner Shimony's less well known---but, perhaps, in the long run equally significant and, possibly, even more visionary intellectual engagement---with the question of the relationship between mind and matter, which only a few physicists of our time have been capable of engaging or prepared to engage seriously because of its more philosophical and, for the time being, less mathematically tractable character. 

As a philosopher, Shimony focused largely on the question of the relationship between knowledge and the world. First, it should be noted that he did not view these two philosophical concerns as distinct from study of the implications of entanglement {\em per se}. Indeed, the most well known phrase he coined which will remain with us is that of ``experimental metaphysics,'' another important usage of his being the transposition from politics to physics of the expression ``peaceful coexistence,'' as pertaining to the relationship between quantum mechanics and special relativity. Shimony long held that any adequate philosophy must take science into account and, moreover, was very confident that the latter has provided a tremendous, genuine increase in human knowledge; he saw the validity of the scientific approach to knowledge as manifestly evident in its results and logic.

As seen from the title of his collection of early papers, {\em The Search for a Naturalistic World View}, Shimony enganged the relationship between the mental and physical via his search for a way to harmonize epistemology and scientific realism. This question is engaged most explicitly in his paper ``Reality, causality, and closing the circle.''
\begin{quote}
I propose to relate the problem of realism to a program which is familiar in systematic philosophies of the past: to understand the knowing subject as an entity in nature and to assess claims to knowledge in the light of this understanding. Such a program aims at the integration of epistemology with the natural sciences and metaphysics. It intends to show how claims to human knowledge of the natural world can be justified, and in turn how the resulting view of the world can account for the cognitive powers of the knowing subject. For brevity I shall refer to this program as ``closing the circle."  
({\it ibid.})
\end{quote}
Shimony's perspective differed from those of many recent philosophers of science who have dealt with this problem as advocates of naturalized epistemology, that is to say those taking the epistemological approach in which the evaluation of the status of human claims to knowledge of the world depends on biology and psychology. He held instead that nature involves an irreducibly mentalistic aspect and did not view the so-called ``first-person" and ``third-person" considerations in epistemology as fully distinct. For him, the closing of the epistemological circle is an ongoing program. 

Shimony rejected the notion that the closing of this circle has been shown impossible; rather, he was optimistic about its prospects. A significant element of this program is the correction of distortions of the objective nature of the world due to the nature of the human subject. Related to the notion that nature involves an irreducibly mental aspect, he believed that there is a continuum from where humans stand, potentially beyond what the human imagination can acheive; he believed that the whole evolutionary perspective requires that humanity has come from something endowed with proto-mentality.\footnote{cf. https://www.aip.org/history-programs/niels-bohr-library/oral-histories/25643 } One possibility, beyond the blanket thesis that there is an irreducible mentalistic aspect of the world, is that there exists direct physico-physical interaction, something which Shimony's teacher Wigner considered might be occurring in relation to quantum-mechanical processes in light of the difficulty in the study of quantum measurement known as the ``measurement problem,'' although  
Shimony also pointed out that ``A warning is needed\ldots against possible misunderstanding of [the term `experimental metaphysics']. One should not anticipate straightforward and decisive resolution of metaphysical disputes by the outcomes of experiments.''\footnote{{\em Quantum Reality, Relativistic Causality, and Closing the Epistemic Circle}.} 
Although his own views as to a possible solution differed from those of his physics doctoral advisor and although his thesis work did not involve the measurement problem, Shimony was very much engaged in discussion with Wigner on the subject at that time. Both were interested in the thinking of Whitehead and the idea that the mental may ultimately be connected with its resolution. Both came to consider the physicalist approach to the mental as misguided and rejected the idea that there is a conservative solution to the quantum measurement problem, one not requiring a fundamental modification of existing physical theory. Finally, both produced results supporting this view.

Shimony saw the quantum measurement problem as a genuine and unsolved one central to the closing of the epistemological circle because the unitary transformation of state according to the Schroedinger evolution, which preserves quantum superpositions, predicts that measurement outcomes do not arise when succesive measurements are made of quantities represented by non-commuting quantum operators. As he put it, ``you don't get data. You have no experimental results. Therefore, you cannot close the circle.''\footnote{\it ibid.} 
However, the closing of the circle was more important for Shimony than a solution of the quantum measurement problem because definite physical outcomes of measurements constitute only a necessary but not a sufficient condition for human experience and inference: ``The greatest obstacle to `closing the circle' is the ancient one which haunted Descartes and Locke---the mind-body problem. Does contemporary microphysics have any implications concerning this problem?'' 

This perspective on the nature of mentality was connected with his interest in the philosophy of Alfred North Whitehead, who approached the mind--body problem similarly to the approach of Leibniz in the monadology and viewed the fundamental things in the natural world as possessing a proto-mental aspect. However, Shimony did not accept this analysis due to the difficulties it has in accomodating evidence against the intrinsic identity of such entities as  electrons, which this picture would endow with unequivocal individuality. 
For his part, he conjectured that a world in which the metaphysical principles emerging from ``experimental metaphysics,'' namely, objective indefiniteness, objective chance and probabilities, and entanglement of physical systems, harmonizes better in the face of apparent dualism than the metaphysics principles of classical physics and could ultimately lead to a resolution of the mind--body problem.

Another thesis rejected by Shimony was that observation must be theory laden. He viewed as strong evidence of the incorrectness of that idea the fact that in their own tests of the Bell inequality, the experimental physicists Clauser and Holt each found results more like the result the {\em other} desired, not the one each hoped for, one being in favor of local hidden variables and one in favor of full agreement with quantum mechanics, respectively, neither of them getting the sort of result personally anticipated. 
Abner Shimony was firmly convinced that physics in his time had made real and substantial progress. However, he also believed that the reconciliation of those aspects of the world that it has uncovered is a daunting task, particularly with regard to the question of how to fully harmonize quantum theory and space-time theory.
He ended his resum\'e, entitled ``Unfinished Work: A bequest,'' of open questions which he himself had pursued over the course of his career, with the following comment regarding this fundamental question. 
``Of all the unfinished work that I am offering you as my bequest, this proposal about physics in the small is the most speculative and difficult. In the long run it probably will be accomplished, but I nevertheless must remind you of Pogo's warning, `We are surrounded by unsurmountable opportunities'.''\footnote{Pogo is a cartoon figure of the socio-political sartirist Walt Kelly. This character is also known for the comment ``We have met the enemy and he is us.''}

\end{document}